\documentclass[amssymb,twocolumn]{revtex4}
\usepackage{graphicx}
\usepackage{dcolumn}
\usepackage{amsmath}
\usepackage{amssymb}

\begin{document}

\title{Macroscopic control parameter for avalanche models for bursty transport.}

\author{S. C. Chapman$^{1,3}$, G. Rowlands$^1$ \& N. W. Watkins$^{2,3}$}

\affiliation{$^{1}$Centre for Fusion, Space and Astrophysics,
University of
 Warwick, UK}
 \affiliation{$^{2}$Physical Sciences Division, British Antarctic Survey (NERC),
 UK}
 \affiliation{$^{3}$also at the Kavli Institute for Theoretical Physics, Santa Barbara, USA}
\date{\today}

\begin{abstract}
Similarity analysis is used to identify the control parameter $R_A$ for the subset of avalanching systems that can exhibit Self- Organized Criticality (SOC). This parameter expresses the ratio of driving to dissipation. The transition
 to SOC, when the number of excited degrees of freedom is maximal, is found to occur when $R_A \rightarrow 0$. This is in the opposite sense to (Kolmogorov) turbulence, thus identifying a deep distinction between turbulence and SOC and suggesting an observable property that could distinguish them.
 A corollary of this similarity analysis is that SOC phenomenology, that is, power law scaling of
   avalanches, can persist for finite $R_A$, with the same $R_A \rightarrow 0$ exponent, if the system supports a sufficiently
    large range of lengthscales; necessary for SOC to be a candidate for physical ($R_A$ finite) systems.
\end{abstract}

\maketitle

\section{Introduction}

It is increasingly recognized that a large group of physical systems can be characterized as  driven, dissipative, out-of-equilibrium and having a conservation law or laws (see the comprehensive treatments of \cite{sornette,sethnabook}). They usually have many degrees of freedom (d.o.f.), or excited modes, and long range correlations leading to scaling or multiscaling. Two examples are fully developed turbulence (see e.g. \cite{frisch,bohr}) and Self- Organized Criticality (SOC, \cite{BTW87,jensenbook,frette}).

The SOC paradigm has found particular resonance with workers attempting to model, and understand, `bursty', scale free transport and energy release in magnetized plasmas (for a recent review, see, for example \cite{rev1}). Simplified avalanche models have been proposed, and points of contact with the data investigated, in the astrophysical context; most notably to describe magnetospheric activity (\cite{mag1,soc3,soc4,soc6}, see also the review \cite{magrev1} and refs. therein), the dynamics of the solar corona (e.g. \cite{soc1,sun1,soc1a,sun2}, see also the review \cite{sunR1}), and accretion disks (e.g. \cite{ac1,ac2,ac3}). In the context of magnetically confined laboratory plasmas there have been extensive efforts to construct avalanche models that make points of contact with the system under study, and to establish signatures characteristic of SOC dynamics in experiments (e.g. \cite{tok3,soc2,tok5,tok7,tok8,tok9,tok10,tok11,soc5,tok13,tok13a,tok14,tok15,tok13b,tokdev1}). There have also been attempts to establish whether the signatures of SOC can emerge from magnetohydrodynamic (MHD) or reduced MHD models, (e.g. in the solar coronal context \cite{sun3,sun4}).

 Since the original suggestion of Bak et al in \cite{BTW88} that SOC ``... could be considered a toy model of generalized turbulence"  there has been continuing debate on the possible
relationship of turbulence to
SOC \cite{turbsoc,bofetta,bramprl,bramN,chapjpa}. Similarities in the statistical signatures of turbulence, and systems in SOC, have been noted (see e.g. \cite{menech,sreeniv}).
In particular, it has recently been argued in the context of astrophysical plasmas that SOC and turbulence  are  aspects of a single underlying physical process  (see \cite{mayaturb,uritskyturb} and references therein).
  However, the extent to which SOC, as opposed to turbulence, uniquely captures the observed dynamics in magnetically confined laboratory plasmas (see \cite{krommes2000,krommes1999,avobs}) or is indeed consistent with it (see \cite{tok16,tok17}) has  been brought into question. Key observables such as power law distributions of patches are not unique to SOC systems (for an example see \cite{chaprobust}, also the comprehensive discussion in \cite{sornette}).

 Our focus here is then to establish the macroscopic similarities and differences between turbulence and SOC in the most general sense.
 A central idea in physics is that complex and otherwise intractable behavior may be quantified by a few measurable macroscopic control parameters.
In fluid turbulence,  the Reynolds number $R_E$ expresses the ratio of driving to dissipation and  parameterizes the transition from laminar to turbulent flow. Control parameters such as the Reynolds number can be obtained from dimensional analysis (see e.g.\cite{frisch,barenblatt}), without reference to the detailed dynamics. From this perspective the level of complexity resulting from the detailed dynamics is simply characterized by the number $N$ of  excited, coupled d.o.f. (or energy carrying modes) in the system. The transition from laminar to turbulent flow then corresponds to an (explosive) increase in $N$. The nature of this transition, the value of the $R_E$ at which it occurs, and the rate at which $N$ grows with $R_E$ all depend on specific system phenomenology. Dimensional arguments, along with the assumptions of steady state and energy conservation, are however, sufficient to give the result that $N$ always grows with $R_E$ (as in  \cite{k41}, see also \cite{frisch}.

We anticipate that an analogous control parameter for complexity, $R_A$, will exist for  the wider group of systems discussed above.  Interestingly it is now known that such a control parameter that expresses the ratio of driving to dissipation
 does indeed exist for SOC. In this paper we will give a prescription to obtain $R_A$  generally from dimensional analysis, that is, without reference to the range of detailed and rich phenomenology that any given system will also exhibit. The rate at which $N$ varies with $R_A$ is again dependent on  this detailed phenomenology. We will see that similarity arguments, along with the assumptions of steady state and energy conservation, are however, sufficient to determine whether or not $N$ grows with $R_A$.

The question of control parameters in Self-Organized Criticality was initially controversial, as the name leads one to expect. It was originally argued (\cite{BTW87,BTW88}, see also \cite{sethna}) that avalanching  systems  self organized to the SOC state without a tuning parameter. Subsequent analysis has
established a consensus (see \cite{vesp,vergeles,dickman,sethnabook,sornette,prucom}) that some
tuning exists, at least in the sense that SOC is a limiting behavior in the driving rate $h$ and the
dissipation rate $\epsilon$, such that $h/\epsilon \rightarrow 0$ with
$h$,$ \epsilon \rightarrow 0$, (and $h \leq \epsilon$, that is, a
steady state).
  This understanding is exemplified in Jensen's  constructive
  definition as given in \cite{jensenbook} of SOC as  the behavior of ``slowly driven interaction dominated thresholded" (SDIDT) systems. Clearly then, $h/\epsilon$ plays the role of a control parameter.
  This SDIDT limit $h/\varepsilon \rightarrow 0$ has been investigated extensively (e.g. \cite{vesp,vergeles}), most recently with respect to finite size scaling in the limit of increasingly large system size (e.g. \cite{pru2006,prucom}).

Here we are concerned with the relevance to SOC to physically realizable systems, and in particular natural ones, where the system size is finite and the driving may be unknown and/or highly variable. Our focus is on parameterizing the level of complexity of the system as we take it away from the SDIDT limit by increasing the driver, in a system of large but fixed size. For avalanche models exhibiting SOC, we will argue that distinct realizable avalanche sizes play the role of excited d.o.f. of the system. The SOC state is then characterized by maximal excited d.o.f., that is, avalanches occurring on all lengthscales supported by the system. Far from the SOC state, the system becomes ordered with few excited d.o.f. and exhibits laminar flow.  The SDIDT
 limit is reached by taking
$R_A$ to zero and we will show that this indeed maximizes the number of excited d.o.f. $N$.
 The SDIDT limit is  thus in the
opposite sense to fluid turbulence which maximizes $N$ at $R_E \rightarrow \infty$.

This suggests a possible means to distinguish observationally between turbulence and SOC in observations and experiments of driven, magnetically confined plasmas.
 For example, a power law region in the power spectral density of some quantity that probes the flow is often identified in both laboratory and astrophysical confined magnetized plasmas (e.g. \cite{tok8,tok9,tok10,tok11}) and is discussed in the context of both SOC and turbulence. This power law region will always be of finite spatio- temporal range (an 'inertial range' of the cascade). Our results imply that this 'inertial range' will decrease as we increase the drive for SOC, whereas it will increase for turbulence- providing an experimental or observational test to distinguish these phenomena.

Our relationship between $R_A$ and $N$ implies the possibility of large but finite $N$ for small but non- zero $R_A$;
hence  an important
corollary is that SOC phenomenology can quite generally persist
under conditions of finite drive in a sufficiently large bandwidth
system. This has been seen in specific avalanche models (see \cite{corral,nickgrl,uritsky}). Here, since
our result flows from dimensional analysis, we will see that this is a generic property of avalanching systems.

\section{Similarity analysis and control parameter}

We shall focus on how the well-established techniques-similarity analysis, as described in \cite{barenblatt}, and the $\Pi$-theorem obtained by Buckingham in \cite{buckingham} nearly a century ago can be used to reveal new information about  avalanche models exhibiting SOC (i.e. \cite{BTW87,frette,jensenbook,sornette,sethna,dickman}).

The  systems that we have in mind all have strongly coupled excited d.o.f. that
transport some quantity from the driving to the
dissipation scale. They have the following properties:

I. The many excited d.o.f. of the system are coupled;  there is
some dynamical quantity that freely flows over all the excited d.o.f. of
the system. We can characterize a flux  $\varepsilon_l$  of this quantity associated with processes that occur on lengthscale $l$, that is, $\varepsilon_l$ is the transfer rate of the dynamical quantity through $l$ to neighboring lengthscales.

II. The system is not necessarily in equilibrium but is in a steady
state on the average.

III. The dynamical quantity is conserved so that given (II) the injection rate $\varepsilon_{inj}$ balances the dissipation rate $\varepsilon_{diss}$, that is   $\varepsilon_{inj}\sim \varepsilon_l\sim \varepsilon_{diss}$ in an ensemble averaged sense.

IV.  The  solution is of a scaling type, that is:
\begin{equation}
N\sim \left(\frac{L_0}{\delta l}\right)^\alpha
\end{equation}
where $\alpha>0$ and $L_0$ and $\delta l$ are the largest and smallest lengthscales respectively that are supported by the system.

V. The number of excited d.o.f. can be parameterized by a single macroscopic control parameter.

We will identify the control parameter for these
systems in terms of known macroscopic variables by  formal dimensional analysis
(similarity analysis or Buckingham $\Pi$ theorem, see e.g.
\cite{buckingham,barenblatt}). The essential idea is that the
system's behavior is captured by a general function $F$ which
only depends on the \emph{relevant} variables $Q_{1..V}$ that
describe the system. Since $F$ is dimensionless it must be a function of the possible dimensionless groupings, the
$\Pi_{1..M}(Q_{1..V})$, which can be formed from the $Q_{1..V}$.
The (unknown) function $F(\Pi_1,\Pi_2,..\Pi_M)$ is universal,
describing all systems that depend on the $Q_{1..V}$ through the
$\Pi_{1..M}(Q_{1..V})$ and the relationships between them. If one
then has additional information about the system, such as a conservation property, the $\Pi_{1..M}(Q_{1..V})$  can be related to each other
to make $F$ explicit. Thus this method can lead to information
about the solution of a class of systems where the governing
equations are unavailable or intractable, often the case
for complex systems where there are a large number ($N$
here) of strongly coupled d.o.f.. If the $V$ macroscopic variables are
expressed in $W$ physical dimensions (i.e. mass, length, time) then there
are $M=V-W$ dimensionless groupings.

The properties I-V above restrict the choice of relevant $Q_{1..V}$.
First, we have only specified there is a transfer rate on lengthscale $l$, $\varepsilon_l$ (property I) of some dynamical quantity, its precise nature is irrelevant.
Consequently,  the only physical dimensions of the transfer rate $\varepsilon_l$ relevant to the problem are length and time, so that
$W=2$. Second, property (V) is that there is a single control
parameter, $\Pi_1$ which  may be
expressed as a function of the number of excited degrees of freedom
$N$. To incorporate the scaling property (IV) we will seek solutions such that $\Pi_2=g(L_0/\delta l)=f(N)$. This means that the system's behaviour is captured by some
$F(\Pi_1,\Pi_2)=C$ which fixes $M=2$ ($C$ is a constant).
The
$\Pi_1$ and $\Pi_2$ are related to each other via properties II and III (conservation and steady state).
We then have that $V=4$; there are always four relevant macroscopic variables to
consider.

 T see this in action, we begin with a relatively well understood example, namely Kolmogorov (K41) turbulence. Our aim here is to straightforwardly illustrate the above approach by obtaining the control parameter, the Reynolds number
 $R_E$ as a function of $N$ via dimensional analysis; for a detailed discussion of the universal scaling properties of K41 turbulence and their origin in the Navier Stokes equations see for example \cite{frisch}. As above, for K41  we have four relevant
macroscopic variables (given in Table 1) and
two dimensionless groups:
\begin{equation}
\Pi_1=\frac{UL_0}{\nu}=R_E,\hskip 2pt
\Pi_2=\frac{L_0}{\eta}=f(N)
\end{equation}
\begin{table}[t]
\begin{center}
\caption{$\Pi$ theorem applied to homogeneous
turbulence.}
\begin{tabular}{ccl}
\hline
Variable      &dimension    & description  \\
\hline
$L_0$&$L$& driving length scale\\
$\eta$&$L$&dissipation length scale\\
$U$&$LT^{-1}$&bulk (driving) flow speed\\
$\nu$&$L^2T^{-1}$&viscosity\\
\hline
\end{tabular}
\label{tab1}
\end{center}
\end{table}
$\Pi_1$ is just the Reynolds number $R_E$ of the flow, and the ratio of lengthscales
$\Pi_2$ is  related to the number of d.o.f. $N$
that can be excited. We now see how $R_E$ is related to $f(N)$ by relating $\Pi_1$ to $\Pi_2$. For
incompressible fluid turbulence, our dynamical quantity $\varepsilon_l$ is the
time rate of energy transfer per unit mass  through
length scale $l$. The procedure is then as follows:
\begin{enumerate}
\item  conservation
and steady state imply (ensemble averaged) that $\varepsilon_{inj}\sim \varepsilon_l\sim \varepsilon_{diss}$; that is, the average energy injection rate $\varepsilon_{inj}$ balances the average energy dissipation rate $\varepsilon_{diss}$.
\item  the rate at which energy is transferred to the fluid is from dimensional analysis:
$\varepsilon_{inj}\sim U^3/L_0$
\item Dimensional analysis of Navier Stokes gives  $\varepsilon_{diss}\sim \nu^3/\eta^4$
\item $\varepsilon_{inj}\sim\varepsilon_{diss}$ then relates $\Pi_1$ to $\Pi_2$:
\begin{equation}
R_E=\frac{UL_0}{\nu}\sim\left(\frac{L_0}{\eta}\right)^\beta
\end{equation}
and fixes exponent $\beta=4/3$
\item the solution is of scaling type, so that:
\begin{equation}
N \sim \left(\frac{L_0}{\eta}\right)^\alpha
\end{equation}
with $\alpha >0$ by definition
\item thus
\begin{equation}
R_E\sim \left(\frac{L_0}{\eta}\right)^\beta \sim N^{\beta_N}
\end{equation} and $\beta_N=\frac{\beta}{\alpha} >0$
\end{enumerate}
The value of the exponents $\alpha$ and $\beta$ will depend on the detailed phenomenology of the turbulent flow. An
 estimate based on K41 for example, with $\beta=4/3$ from the above, and $\alpha=D=3$ where $D$ is Euclidean dimension\cite{frisch}, implies a high degree of disorganization and will be modified for example if the turbulence is intermittent.
 Importantly, the only property of turbulence with which we are concerned here is that
  both $\beta >0$ and  $\alpha >0$ so that $\beta_N=\beta/\alpha >0$.
This identifies the  Reynolds number as the
control parameter for a process (turbulence) which
simply excites more active modes or d.o.f. as we increase $R_E$.

We now see how above arguments apply to other systems as defined above, in particular to avalanche models.
Without recourse to details of the system,
 similarity analysis will be sufficient to obtain the relationship between the control parameter $R$ and
the number of degrees of freedom $N$ of the form:
\begin{equation}
R \sim N^{\beta_N}
\end{equation}
 The value of the exponent $\beta_N$ will depend on the details of these systems but crucially we will see that
the sign of $\beta_N$ is fixed by the similarity analysis. This is sufficient to establish whether or not, as in the case of turbulence, increasing $R$ increases the
the number of excited d.o.f. in the system.

\section{Control parameter for avalanching systems}

 We now envisage a generic avalanche model in a system of size
$L_0$ where the height of sand is specified on a grid, with nodes
at spacing $\delta l$. Sand is added to individual nodes, that is,
on length scale $\delta l$  at an average time rate
$\varepsilon_{inj}=h$ per node. There is some process, here
avalanches, which then transports this dynamical quantity (the
sand) though structures on intermediate length scales $\delta l
<l<L_0$. Sand is then
  lost to the system (dissipated) at a time rate $\epsilon$ over the
  system size $L_0$.
   On  intermediate length scales $\delta l <l<L_0$, sand is conservatively
transported via avalanches (see also \cite{decarvalho,decar2,decar3}). In our discussion here we follow \cite{BTW87} and assume that the transport timescale is fast, so that avalanches occur instantaneously and do not overlap. There must be some detail of the
internal evolution of the pile that maximizes the number of length
scales $l$ on which avalanches can occur. For avalanche models this is
the property that transport can only occur locally if some local
critical gradient is exceeded; as a consequence the pile evolves
through many metastable states.  If these length scales represent excited
d.o.f. then  the number $N$ of
d.o.f. available  will be bounded by $L_0$ and $\delta l$ so that
$N\sim (L_0/\delta l)^\alpha$, with $D \geq \alpha\geq 0$ for
$D>1$ ($\alpha$ may be fractional).

The four relevant variables for the avalanching system are given in
Table 2. The two dimensionless
groups are:
\begin{equation}
\Pi_1=\frac{h}{\epsilon}=R_A,\hskip 2pt
\Pi_2=\frac{L_0}{\delta l}=f(N)
\end{equation}

\begin{table}[t]
\begin{center}
\caption{$\Pi$ theorem applied to an avalanching
system. The sand carries a property  with
dimension $S$.}
\begin{tabular}{ccl}
\hline
Variable        &dimension    & description  \\
\hline
$L_0$&$L$&system size\\
$\delta l$&$L$&grid size\\
$\epsilon$&$ST^{-1}$& system average dissipation/loss rate\\
$h$&$ST^{-1}$&average driving rate per node\\
\hline
\end{tabular}
\label{tab1}
\end{center}
\end{table}

We will now relate the control parameter $\Pi_1=h/\epsilon$  to the number of excited d.o.f.
by following the same procedure as above. $\varepsilon_l$ now refers to the time rate of transfer of 'sand'   through
length scale $l$.
\begin{enumerate}
\item  conservation
and steady state imply (ensemble averaged) $\varepsilon_{inj}\sim \varepsilon_l\sim \varepsilon_{diss}$
\item In Euclidean dimension $D$ there are $(L_0/\delta l)^D$
nodes; $D>0$ by definition. The rate at which 'sand' is transferred to the pile is then from dimensional analysis:
$\varepsilon_{inj}\sim h(L_0/ \delta l)^D$
\item the system average dissipation rate is defined as $\epsilon=\varepsilon_{diss}$
\item $\varepsilon_{inj}\sim\varepsilon_{diss}$ then gives
$h(L_0/ \delta l)^D \sim \epsilon$ or:
 \begin{equation}
R_A=\frac{h}{\epsilon}\sim \left(\frac{\delta l}{L_0}\right)^D
\end{equation}
thus in the above notation fixes $\beta=-D<0$
\item
 the number $N$ of
d.o.f. available  will be bounded by $L_0$ and $\delta l$ so that:
\begin{equation}
N\sim \left(\frac{L_0}{\delta l}\right)^\alpha
\end{equation}
with $D \geq \alpha\geq 0$ for
$D>1$ (the value of $\alpha$ depends on the details and may be fractional).
\item thus
\begin{equation}
R_A=\frac{h}{\epsilon}\sim \left(\frac{\delta l}{L_0}\right)^D\sim
N^{-\alpha D}\sim N^{\beta_N}
\end{equation}
and $\beta_N=\frac{\beta}{\alpha}=-\frac{D}{\alpha} <0$
\end{enumerate}
We then have that  the number of excited d.o.f.
\emph{decreases} as we increase the control parameter
$R_A=h/\epsilon$. Thus we recover the SDIDT limit for SOC, namely
$R_A \rightarrow 0$, but now explicitly identify this limit with
maximizing the number of excited d.o.f.. Our result from dimensional analysis
is to obtain $R_A \sim N^{\beta_N}$ and to show quite generally that for the avalanching system $\beta_N<0$.

Our dimensional analysis for the avalanche model maps onto that for K41 turbulence, so in that sense $R_A \equiv R_E$, that is, $R_A$ is the avalanching
 system's 'effective Reynolds number', which expresses the ratio of driving to dissipation. Both $R_E$ and $R_A$   increase with driving of the system, but the system's response is quite different. In the case of K41 turbulence, the system can excite more modes or degrees of freedom and the flow becomes more disorganized, whereas in the avalanche models, less d.o.f. are available so the system is pushed toward order. The essential difference between the two systems in this context is as follows. As we increase the driving in K41 turbulence,
 the smallest lengthscale $\eta$ can decrease (via Navier Stokes) to provide the necessary dissipation to maintain a steady state, and since we have assumed scaling the system simply excites more modes or d.o.f.. On the other hand,  in the avalanche models both the smallest and largest lengthscales are fixed; increasing the driving will ultimately introduce sand at a rate that exceeds the rate at which sand can be transported by the smallest avalanches, as we discuss next.

\section{SOC- like behaviour under intermediate drive}
For avalanching to be the dominant mode of transport of sand, there are conditions
on the microscopic details of the system; specifically, there must
be a separation of timescales such that the relaxation time for the
avalanches must be short compared to  the time taken for the
driving to accumulate sufficient sand locally to trigger an avalanche.
Avalanches are triggered when a critical value for the local gradient is exceeded,
the critical
gradient can be a random variable but provided it has a defined
average value $g$, we have that on average, we would need to add $g \delta l$ sand to a single cell of an initially flat pile to trigger redistribution of sand. The number of timesteps that this would take to occur would on average be $(g \delta l)/(h \delta t)$ where again $\delta l$ is the cell size and $\delta t$ is
the timestep. This gives the condition for avalanching to dominate transport on all lengthscales in the grid $[\delta l, L_0]$, so that avalanches only occur after many grains of sand
have been added to any given cell in the pile and is the strict
SDIDT\cite{vesp,vergeles}  limit:
\begin{equation}
h \delta t \ll g \delta l
\end{equation}
We will now consider an intermediate behaviour (see also \cite{nickgrl})
\begin{equation}
g \delta l<h \delta t \ll g \delta l \left(\frac{L_0}{\delta l}\right)^D
\end{equation}
where the driver is large
enough to swamp of order $h\delta t/(g \delta l)$ cells of the pile at each timestep (each addition of sand),
 but this is still much smaller than the largest avalanches that the system is able to support since the largest possible avalanche in a system of Euclidean dimension $D$ is $\left(L_0/\delta l\right)^D$ cells.

 For a given physical realization of the sandpile, that is, fixed box size
$L_0$ and grid size $\delta l$, successively increasing $h \delta t $ above $ g \delta l$ then
successively increases the smallest avalanche size (to some $\delta l' > \delta l$).
Ultimately as $h$
and hence $R_A$ is increased to the point where $h\delta t \sim g \delta l
\left(L_0/\delta l\right)^D$  there will be a crossover to laminar
flow, as each addition of sand drives avalanches that are on the size of the system.

We now assume that the avalanching process is self- similar, so that the system is large enough that  the probability density of avalanche sizes $S$ is $P(S) \sim S^{-\gamma}$ over a large range of $S$; that is, finite sized effects do not dominate. Consequently
this intermediate, finite  $R_A$ behavior will be
 `SOC like', with avalanches occurring within the range of lengthscales $[\delta l', L_0]$ with power law statistics sharing the same exponent $\gamma$ as at the SDIDT limit.

  We will illustrate these remarks with simulations of the Bak, Tang and Wiesenfeld (BTW) (see \cite{BTW87}) sandpile in 2D, where the
driving occurs randomly in time and is spatially restricted to the
`top' of the pile. In all cases shown, the critical gradient (threshold for avalanching) is $g =4\delta l^{-1}$, and normalized distributions of the number of topplings in an avalanche  $S$ are shown (we take topplings as a measure of avalanche size following \cite{BTW87}).
 In Figure 1  we plot the results from two simulations in a box of size $L_0/\delta l =100$, under driving rates $h  =4 \delta t^{-1}$ and $h' =16 \delta t^{-1}$. We can see that as we increase the driving rate from $h =4 \delta t^{-1} $ to $h'=16 \delta t^{-1}$ ,  the occurrence probability of the smallest avalanches is reduced,  and on these normalized histograms, the probability of larger events is increased. These larger events for the run with $h'=16 \delta t^{-1}$, that is, for $S\sim [10^2-10^{3.5}]$, still follow the same power law scaling as the $h=4 \delta t^{-1}$ run
 (the precise location on the plot of the crossover in behaviour will depend on details of the dynamics of the pile).
 This is to be anticipated
 provided that transport on these intermediate scales is still dominated by avalanching, that is, intermediate scale avalanches still have the property that they relax on a timescale that is much faster than that required by the driving to initiate an avalanche. If this is the case, then the phenomenology of these intermediate scale avalanches is unchanged by the increase in the driving rate and as a consequence, except close to the crossover in statistics, their scaling exponent is, as we see, unchanged. As the system has self
similar spatial
 scaling  we can also anticipate
obtaining the same solution for these avalanches subject to a rescaling; $S$, which is a measure of avalanche size, will simply scale with $h \delta t$, the  sand which must be redistributed at each timestep since $h \delta t > g \delta l$.
This is shown in the lower plot of Figure 1 where we have rescaled the $h'=16 \delta t^{-1}$ intermediate range driving results by $S \rightarrow S/16$. We can see that power law  regions of the plots that both correspond to avalanching now coincide.

We can go further and anticipate that two realizations of the system, one with
$h$ and $L_0$  and the other with $h'= A h$ and  $L_0'= A L_0$
give the same solution for $P(S)$ under rescaling  $S \rightarrow S'/A^D$.
This is shown in Figure 2 where we compare  two runs of the sandpile (i) with $h  =4 \delta t^{-1}$ and $L_0=100 \delta l$ and (ii) with  $h'=16$ and $L_0'=400 \delta l$, in the same format as Figure 1.  We can indeed see a close correspondence of the avalanche statistics in the power law region of the plot once we have rescaled the $h'=16 \delta t^{-1}$ and $L_0'=400 \delta l$ run by $S \rightarrow S/16$ (at the largest $S$, the histograms do not precisely collapse under this self affine scaling, see \cite{BTW88} for a discussion of the finite size type of scaling properties of the model).

\begin{figure}
\includegraphics[width=0.3 \textwidth]{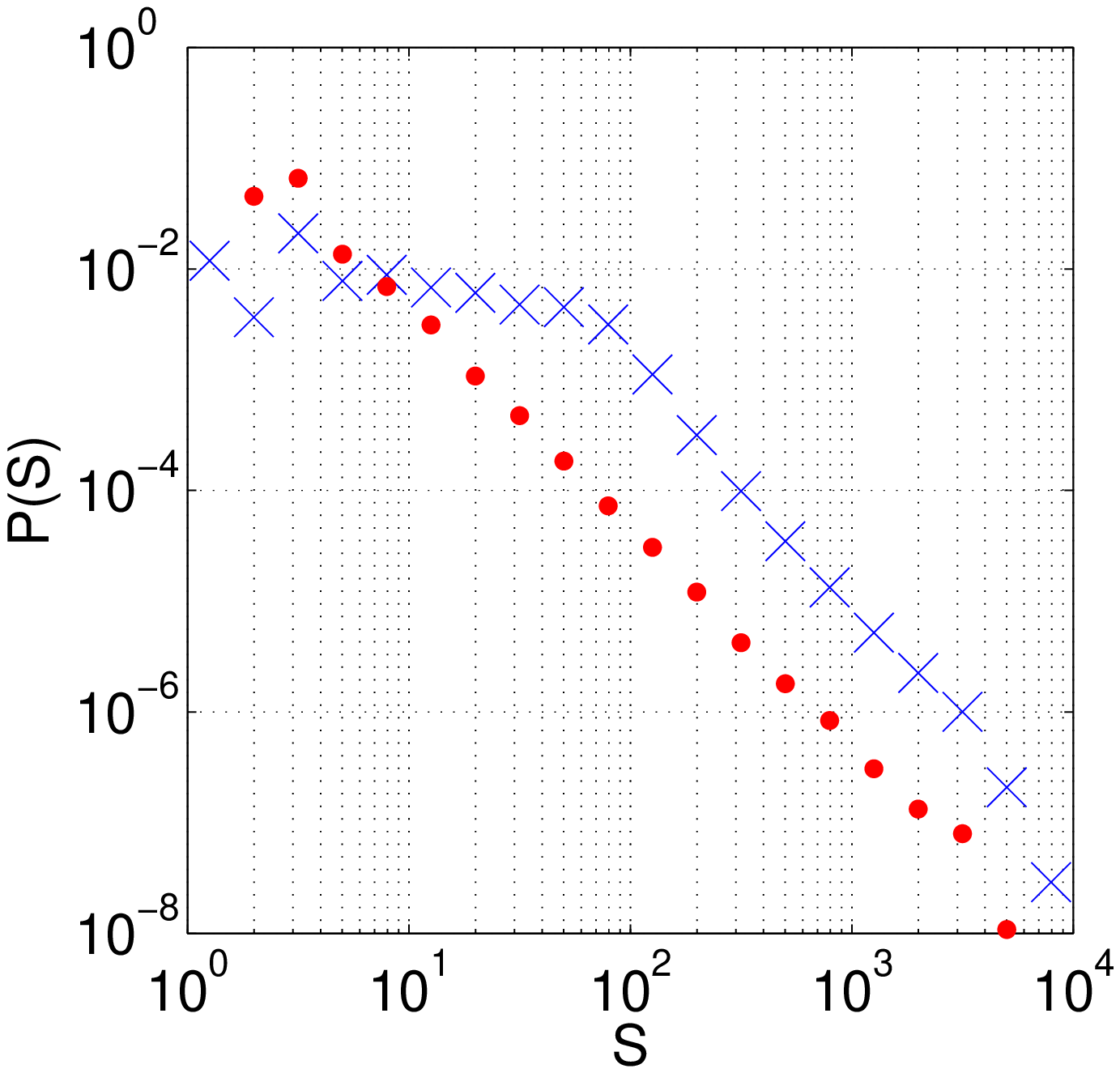}
\hskip 3cm
\includegraphics[width=0.3 \textwidth]{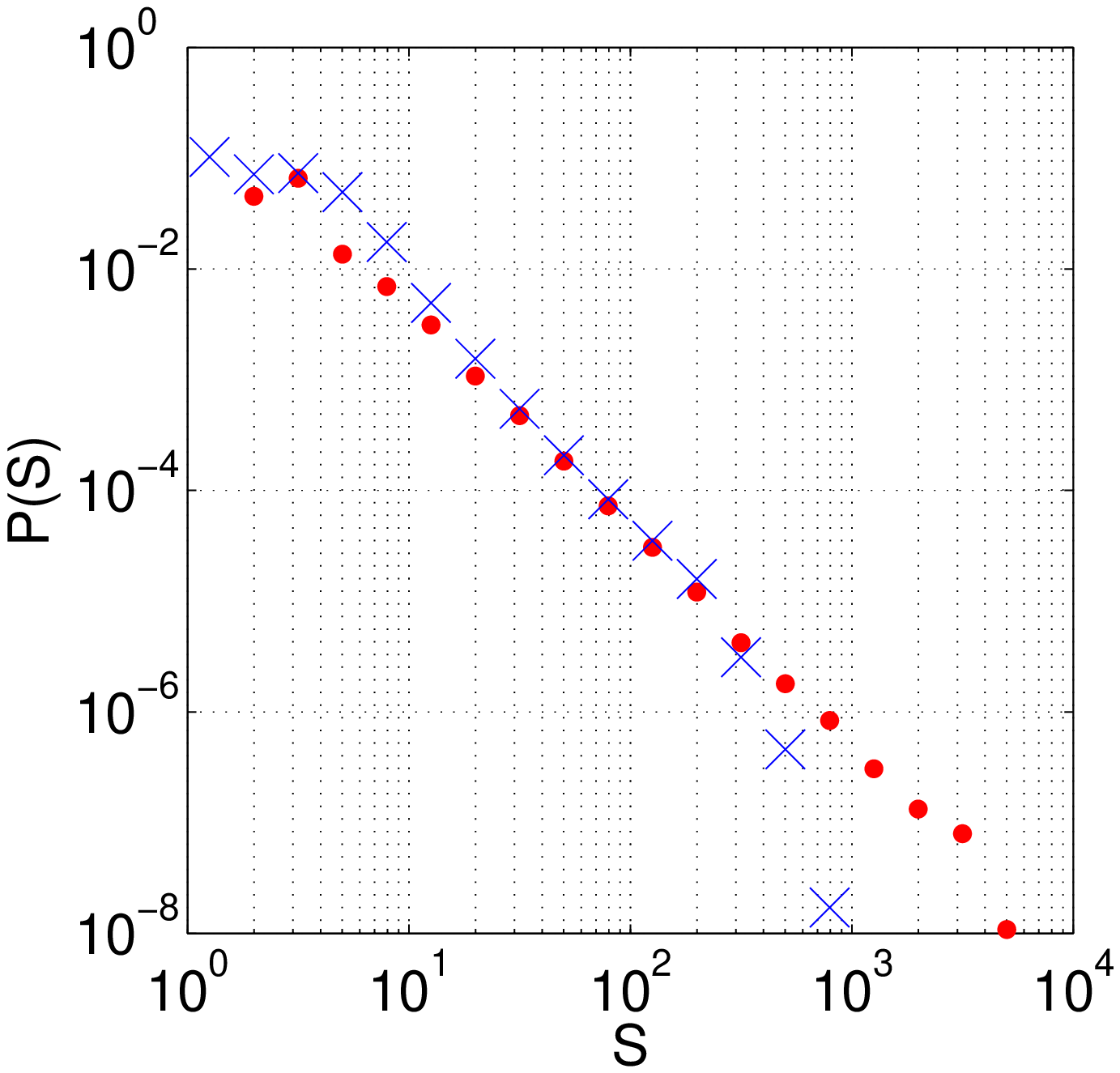}
\caption{(Color online) Avalanche size normalized distributions for two runs of the 2D
BTW\cite{BTW87,BTW88} sandpile driven at the 'top' corner formed
by two adjacent closed boundaries, the other boundaries are open.
$L_0/\delta l=100$ and $h\delta t=4$ ($\bullet$) and $h\delta t=16$ ($\times$); (left)
probability densities; (right) as (a) with probability density for the
$h\delta t=16$ avalanche sizes rescaled $S\rightarrow S/16$.}
\end{figure}
\begin{figure}
\centering
\includegraphics[width=0.3 \textwidth]{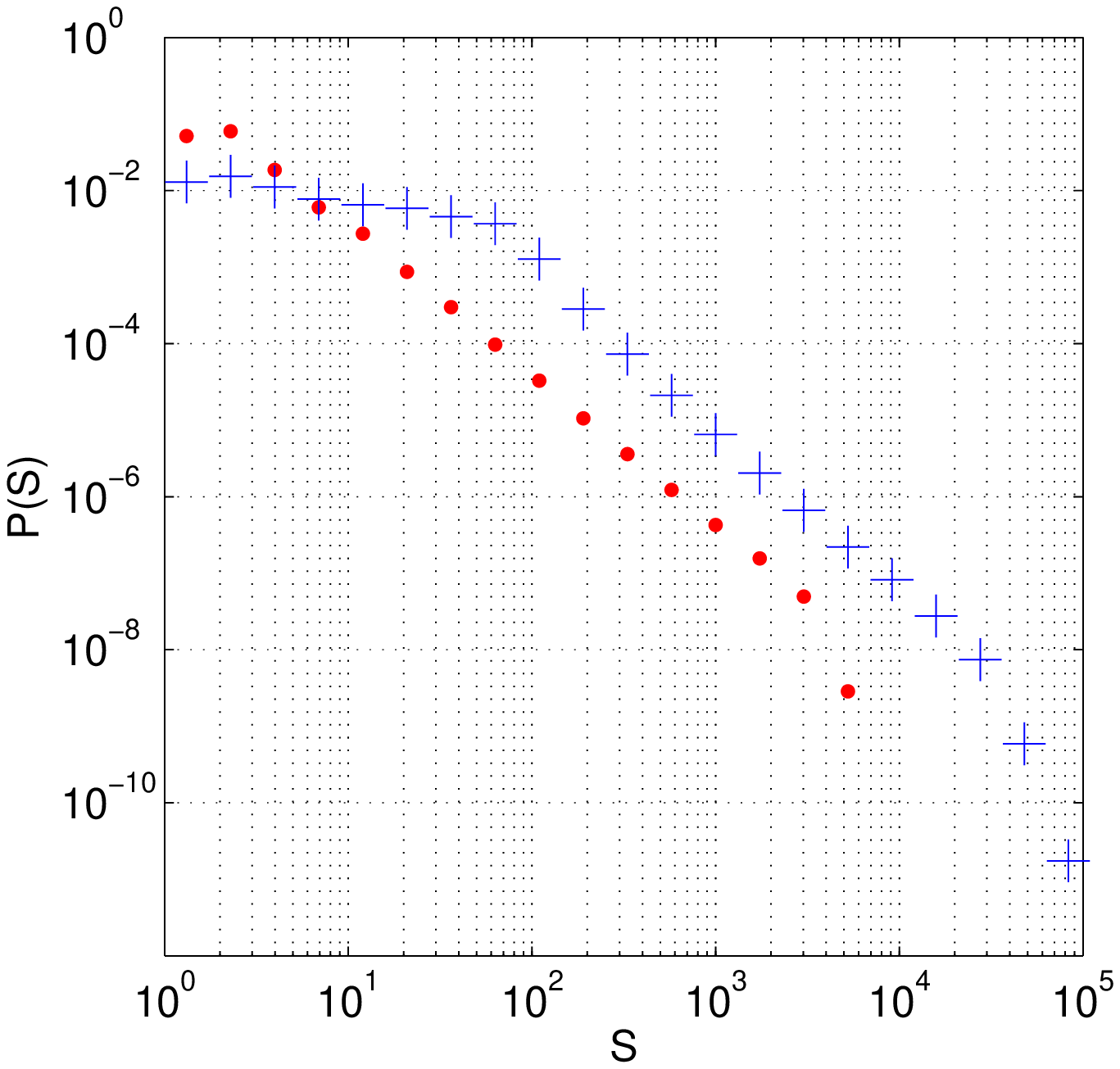}
\hskip 3cm
\includegraphics[width=0.3 \textwidth]{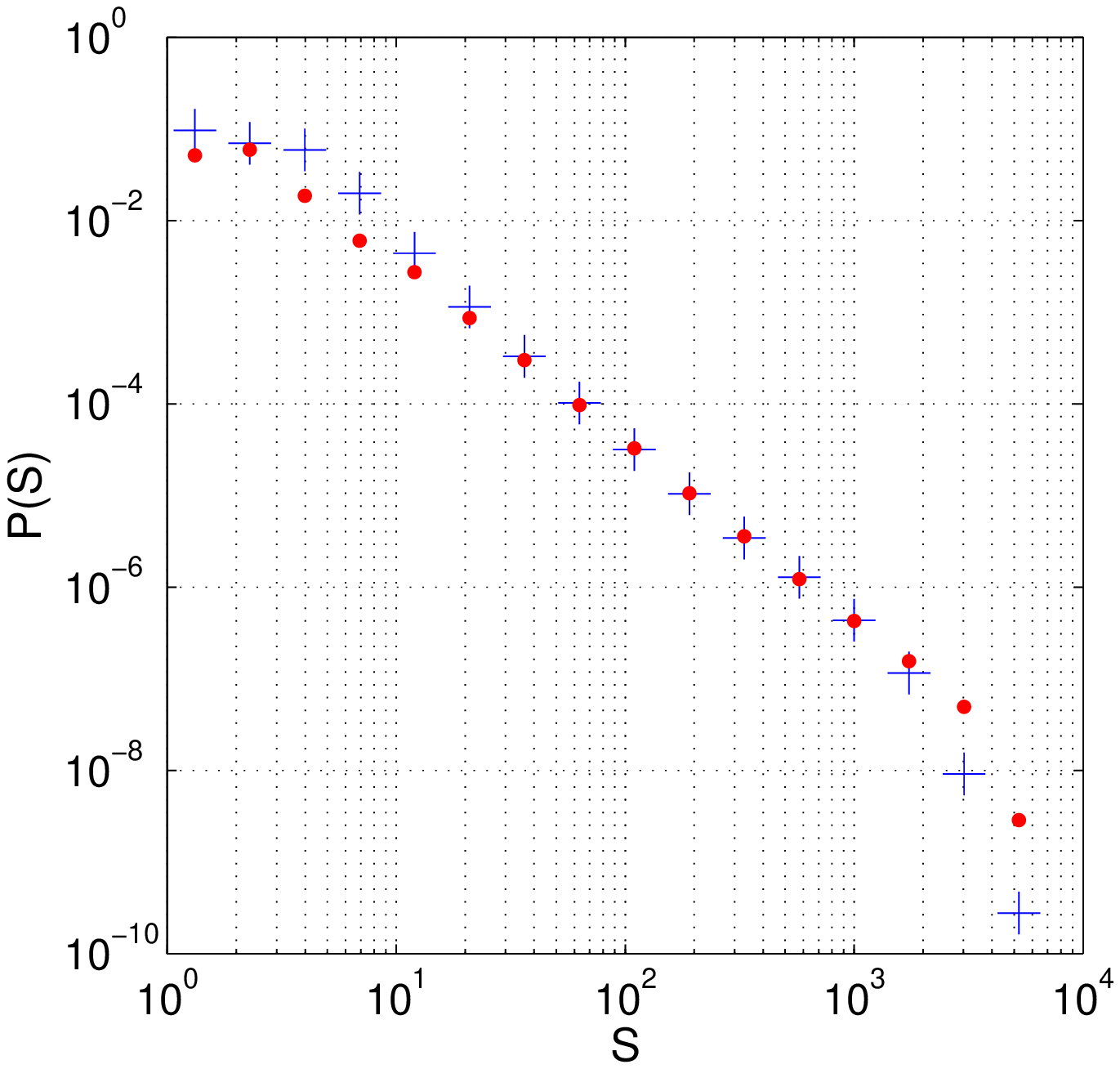}
\caption{(Color online) Avalanche size normalized distributions for
$L_0 /\delta l=100$, $h\delta t=4$ ($\bullet$) and $\frac{L_0}{\delta l}=400$, $h\delta t=16$ ($+$); (left)
probability densities; (right) as (a) with probability density for the
$h\delta t=16$ avalanche sizes rescaled $S\rightarrow S/16$.}
\end{figure}

This establishes a general property of avalanching systems  that
has been seen in several representative SOC
models, such as in \cite{corral,nickgrl,uritsky}. Depending on the
details, specifically, provided that a separation of timescales for avalanching can be maintained, some SOC systems will show scaling in systems where the
drive is in fact highly variable. One could argue that such
robustness against fluctuations in the driving is necessary for SOC
to provide a `working model' in real physical systems where the
idealized SDIDT limit may not be realized.

\section{Conclusions}
We have used similarity and dimensional analysis to discuss high dimensional, driven, dissipating, out of equilibrium systems, in particular avalanching systems that exhibit bursty transport that can be in an SOC state. These  act to transport a dynamical quantity (e.g. for the avalanche models, sand) from the driving to the dissipation scale, in a manner that is conservative, that is steady state on the average, and that shows scaling. The generic nature of this method of analysis implies that our results are not restricted to sandpile models per se, and have wider application to physical systems that show bursty transport, and scaling.
We have postulated that a 'class' of these systems have a single control parameter $R$ which expresses the ratio of the driving to the dissipation and which can be related to the number of excited degrees of freedom $N$. Dimensional analysis then leads to a relationship of the form $R \sim N^{\beta_N}$ and without reference to any detailed phenomenology of the system, determines the sign of $\beta_{N}$.

We have focussed on avalanche models that can exhibit SOC, for which the above identifies the control parameter $R_A=h/\epsilon$. The limit $R_A \rightarrow 0$ is just the well known SDIDT limit of SOC.  Specific avalanching systems will have different values of $\beta_{N}$ but will all share the essential property that we obtain here, that $\beta_{N}<0$ so that that $N$ is maximal under the limit of vanishing driving.
Our formalism for SOC has close correspondence with that for Kolmogorov homogeneous isotropic turbulence. A minimalist interpretation of our results is that Kolmogorov turbulence maximizes the number of excited d.o.f. $N$ under maximal (infinite) driving in contrast to SOC. A maximalist interpretation is that
 $R_A$ is analogous to the Reynolds number $R_E$.  This establishes an essential distinction between turbulence and SOC. Practically speaking, it can for example arise because if we fix the outer, driving scale in Kolmogorov turbulence, the dissipation scale can simply adjust as we increase the driving. Since the system shows scaling, this acts to increase the available degrees of freedom. Avalanching on the other hand, is realized in a finite sized domain (box) and driven on a fixed, smallest scale, so increasing the driving beyond a certain point simply swamps the smallest spatial scales, thus reducing the available degrees of freedom. Increasing the driving then pushes Kolmogorov turbulence toward increasingly disorganized flow, and avalanching systems toward more ordered (laminar) flow.

 A corollary is that SOC phenomenology, that is, power law scaling of
   avalanches, can persist for finite $R_A$ with the same exponent that is seen at the $R_A \rightarrow 0$ limit, provided the system supports a sufficiently
    large range of lengthscales. This has been seen previously for specific realizations of avalanche models \cite{nickgrl} but is shown here to be quite generic; and is
    a necessary property for SOC to be a candidate for physical ($R_A$ finite) systems.
    As the driving is increased, the excited number of degrees of freedom (modes) decreases for SOC and increases for turbulence, so that in principle one could distinguish SOC from turbulence observationally by testing how the bandwidth (range of spatio- temporal scales) over which scaling is observed, varies with the driving rate.

\begin{acknowledgments}
We thank  M. P. Freeman and K. Rypdal for discussions. This research was supported in part by
 the STFC, the EPSRC, and the NSF (under grant No. NSF PHY05-51164).
\end{acknowledgments}

\end{document}